\documentclass[letterpaper,twocolumn,prl,aps,showpacs,superscriptaddress,floatfix]{revtex4-1}
\usepackage[latin1]{inputenc}
\usepackage{bm}
\usepackage[usenames]{color}
\usepackage{multirow}
\usepackage{amssymb}
\usepackage{amsbsy}
\usepackage{amsmath}
\usepackage{stmaryrd}
\usepackage{graphicx}
\usepackage{epsfig}
\usepackage{placeins}
\usepackage{ulem}
\makeatletter
\begin{document}

\title{Spin Transport in the XXZ Chain at Finite Temperature and Momentum}

\author{Robin Steinigeweg}
\email{r.steinigeweg@tu-bs.de}
\affiliation{Institute for Theoretical Physics, Technical University
Braunschweig, D-38106 Braunschweig, Germany}

\author{Wolfram Brenig}
\affiliation{Institute for Theoretical Physics, Technical University
Braunschweig, D-38106 Braunschweig, Germany}

\date{\today}

\begin{abstract}
We investigate the role of momentum for the transport of
magnetization in the spin-$1/2$ Heisenberg chain above the isotropic
point at finite temperature and momentum. Using numerical and
analytical approaches, we analyze the autocorrelations of density
and current and observe a finite region of the Brillouin zone with
diffusive dynamics below a cut-off momentum, and a diffusion
constant independent of momentum and time, which scales inversely
with anisotropy. Lowering the temperature over a wide range,
starting from infinity, the diffusion constant is found to increase
strongly while the cut-off momentum for diffusion decreases.
Above the cut-off momentum diffusion breaks down completely.
\end{abstract}

\pacs{05.60.Gg, 71.27.+a, 75.10.Jm}

\maketitle

Understanding spin transport in quantum many-particle systems is a
fundamental challenge to physics, of immediate relevance to future
information technologies \cite{Wolf2001}, and intimately related to
timely issues of dynamics and thermalization in a more broader context
\cite{Cazalilla2010}. While conventional spin conductors like
silicon \cite{Appelbaum2007}, III-V semiconductors \cite{Stern2008}, carbon
nanotubes \cite{Kuemmeth2008}, or graphene \cite{Tombros2007} necessarily
feature spins which are associated with itinerant charge carriers,
insulating quantum magnets may open new perspectives for spin transport,
with pure magnetization currents flowing solely by virtue of exchange
interactions. Magnetic transport in one-dimensional (1D) quantum magnets
has experienced an upsurge of interest in the last decade due to the
discovery of very large magnetic {\it heat} conduction \cite{sologubenko00}
with mean free paths above $1 \mu m$ \cite{Hlubek2010}. Genuine {\it spin}
transport in quantum magnets remains yet to be observed experimentally,
however long nuclear magnetic relaxation times \cite{thurber01} have been
established, which even allow for manipulation with magnetic fields
\cite{kuehne10}.

Theoretically, significant attention has been devoted to spin
transport in 1D quantum magnets, see Refs.~\onlinecite{zotos-review,
hm07a} for reviews. The dissipation of spin currents is a key issue
in this context and has been analyzed extensively at zero momentum
and frequency in connection with the spin Drude weight
\cite{DrudeWeight}. Spin current dynamics at {\it finite} momentum
remains one of the open questions. In this Letter, we will address
this question for the antiferromagnetic and anisotropic spin-$1/2$
Heisenberg (XXZ) chain
\begin{equation}
H = J \sum_r^N (S_r^x S_{r+1}^x + S_r^y S_{r+1}^y + \Delta \, S_r^z
S_{r+1}^z) \, , \label{H}
\end{equation}
where $S_r^i$ ($i = x,y,z$) are the components of spin-$1/2$
operators at site $r$, $N$ denotes the number of sites, $J > 0$
represents the exchange coupling constant, and $\Delta$ is the
anisotropy. The XXZ chain is a fundamental model to describe
magnetic properties of interacting electrons. It is relevant to the
physics of low-dimensional quantum magnets \cite{johnston00},
ultra-cold atoms \cite{trotzky08}, nanostructures
\cite{gambardella06}, and -- seemingly unrelated -- fields such as
string theory \cite{kruczenski04} and quantum Hall systems
\cite{kim96}.

Early analysis of the time-dependent correlation function of the
local spin density has been performed in the high-temperature limit,
$T=\infty$, suggesting the absence of spin diffusion for $0
\leqslant \Delta \leqslant 1$ \cite{Fabricius1998}. Subsequent,
studies have concentrated on the spin Drude weight at zero momentum
$q=0$ \cite{DrudeWeight}, allowing for no conclusions on diffusion
laws at finite momentum. First low-temperature quantum Monte-Carlo
studies at $q \neq 0$ \cite{alvarez02} found no evidence for spin
diffusion; however, more recent results from bosonization and
transfer-matrix renormalization group \cite{sirker2009} as well as
quantum Monte-Carlo \cite{grossjohann2010} are consistent with
finite-frequency spin diffusion in the small-momentum regime, at
$\Delta=1$ and for low temperatures $T \ll J$, with a spin
diffusion constant $D$ which diverges $\propto 1/T \ln T$. The
physics at intermediate temperatures and arbitrary momenta remains
undisclosed.

Therefore, in this Letter, we consider the transport of
magnetization by analyzing autocorrelations of spin density and
current at finite momenta, covering the complete Brillouin zone, and
at intermediate temperatures $0.5 J \leqslant T \leqslant \infty$
($\hbar = k_B = 1$). We focus on the case of finite anisotropy
$\Delta > 1$, where Eq.~(\ref{H}) features a gapped ground state.
Using a combination of exact diagonalization and perturbation
theory, we uncover a regime of diffusive transport below a finite
critical momentum $q_D$. In this regime, density modes at fixed
momentum $q$ decay with a diffusion constant $D_q$ and our analysis
is consistent with $D_q$ {\it independent} of momentum and inversely
proportional to the anisotropy. As the temperature is lowered from
$T = \infty$, we observe a decrease of the critical momentum and an
almost exponential increase of the diffusion constant. We provide
evidence for a complete breakdown of diffusion above the critical
momentum.

We begin by introducing the generalized diffusion coefficient as a
quantity suitable to describe the evolution of a harmonic spin
density profile close to equilibrium, i.e., in the linear response
regime. To this end, the central quantities we
analyze are the autocorrelation functions $C_{S,q}(t) = \mathrm{Re} \langle
S^z_q(t) S^z_{-q} \rangle/N$ and $C_{J,q}(t) = \mathrm{Re} \langle J^z_q(t)
J^z_{-q} \rangle/N $ of the spin density $S^z_q = \sum_r e^{\imath q r}
S_r^z$ and the spin current $J^z_q = J \sum_r e^{\imath q r} (S_r^x
S_{r+1}^y - S_r^y S_{r+1}^x)$ at momentum $q = 2 \pi k/N$ \cite{mahan2000},
where $\mathrm{Re}$ indicates the real part, $\langle \ldots \rangle$
denotes the canonical equilibrium average at the inverse temperature $\beta
= 1/T$, and $t$ represents the time. Since the density $S^z_q$ and
the current $J^z_q$ are connected by the lattice continuity equation
$\partial_t \, S^z_q = (1-e^{\imath q}) J^z_q$, the autocorrelations are
related by $\partial_t^2 \, C_{S,q}(t) = -\tilde{q}^2 \, C_{J,q}(t)$ with
the abbreviation $\tilde{q}^2 = 2(1-\cos q)$. The generalized, time- and
momentum-dependent diffusion coefficient is {\it defined} via
\begin{equation}
D_q(t) = \frac{\partial_t \, C_{S,q}(t)}{-\tilde{q}^2 \, C_{S,q}(t)}
= \frac{I_q^1(t)}{C_{S,q}(0) - \tilde{q}^2 \, I_q^2(t)} \, .
\label{D}
\end{equation}
To arrive at the right-hand expression in Eq.~(\ref{D}), we integrate the
continuity equation twice, using $\partial_t \, C_{S,q}(t)|_{t=0}=0$ and
introducing the two integrals $I^1_q(t) = \int_0^t \mathrm{d}t' \,
C_{J,q}(t')$ and $I^2_q(t) = \int_0^t \mathrm{d}t' \, I^1_q(t')$.

The left-hand expression in Eq.~(\ref{D}) identifies the quantity
$\tilde{q}^2 D_q(t)$ with the instantaneous decay rate, at time $t$, of
a spin density profile with wave vector $q$ close to equilibrium. Fick's
law corresponds to the case of $D_q(t)= \mathrm{const}$. The main goal of
this Letter is to analyze the time- and momentum-dependence of this quantity
versus temperature. We emphasize that a complete knowledge of this dependence
allows to propagate arbitrarily shaped spin density profiles in time. This
does not only share a common interest with time-dependent density-matrix
renormalization group studies \cite{langer2009}, yet confined to zero
temperature, but even more so may be of relevance to laser pulse induced
time-dependent transport measurements, including recently proposed
time-of-flight and thermal imaging techniques \cite{otter2009}.

\begin{figure}[tb]
\includegraphics[width=0.65\columnwidth]{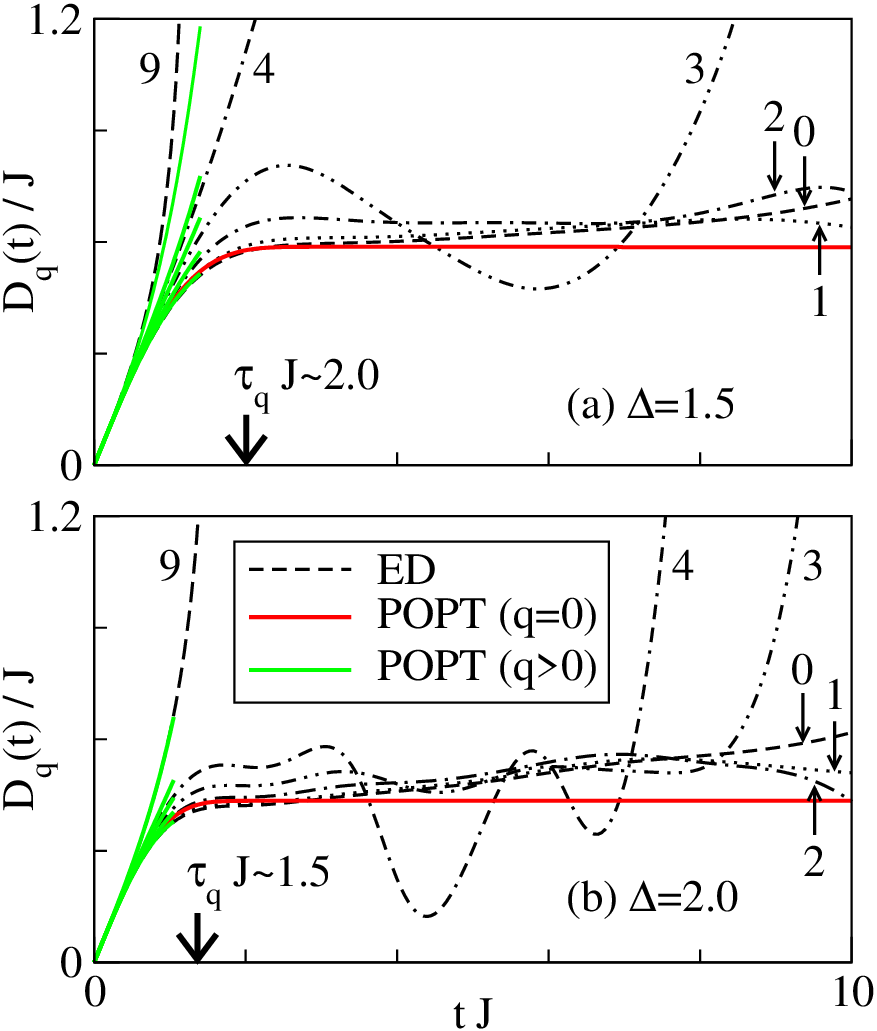}
\caption{(color online) The time- and momentum-dependent diffusion
coefficient $D_q(t)$ at $\beta = 0$ and (a) $\Delta = 1.5$, (b)
$\Delta = 2.0$. ED results are shown for $N= 18$ and $q/(2 \pi/N) =
0,1,2,3,4$, and $9$ (non-solid curves). POPT results are shown for
$q = 0$ (red/dark-colored, solid curves) and $q
> 0$ (green/light-colored, solid curves). Thick arrows on the
$tJ$-axis mark the locations of the current decay time $\tau_q$.}
\label{Fig1}
\end{figure}

Qualitatively, the variation of $D_q(t)$ versus $t$ can be understood from
a standard relaxation-time approximation, in which the current
autocorrelation $C_{J,q}(t)=\exp(-t/t_q) \, C_{J,q}(0)$ decays
exponentially. For short times, $t \ll t_q$, Eq.~(\ref{D}) then yields
$D_q(t)\sim 1-e^{-t/t_q}$, which starts with a linear increase, $D_q(t)
\propto t$, and turns into a `plateau' $D_q(t) \approx \mathrm{const.}$,
starting at $t = \tau_q \gtrsim t_q$. This plateau marks the hydrodynamic
regime. Namely, proceeding to the long-time limit, i.e.~for $t \gg t_q$,
and to the long-wavelength limit, i.e.~for $\tilde{q}^2 (t-t_q) D_q \ll 1$,
Eq.~(\ref{D}) leads to a time-{\it independent} diffusion constant $D_q(t)
= D_0 + {\cal O}(\tilde{q}^2)$, where $D_q = t_q \, C_{J,q}(0)
/C_{S,q}(0)$, which is equivalent to Einstein's relation \cite{ER},
and $D_0 = D_{q=0}$. In principle, partial conservation of currents at
$q=0$, i.e.~the impact of a finite Drude weight at zero frequency
\cite{DrudeWeight}, can also be included into this qualitative
picture. For that case the exponential decay of $C_{J,0}(t)$ has to be
leveled off into $C_{J,0}(t\rightarrow\infty)= \mathrm{const.} > 0$. This
leads to a linear increase $D_0(t \rightarrow\infty) \propto
t$. However, the Drude weight will not be an issue in this Letter. In
fact, there is no zero-frequency contribution of currents at $q \neq 0$,
which follows directly from the continuity equation.

While the gross feature of the preceding relaxation-time {\it ansatz}
can serve as a guideline to interpret the results of unbiased exact
diagonalization data, on which we will report later, it is not justified
{\it a priori}. Therefore, and to gain a deeper insight into the
high-temperature current dynamics generated by the Heisenberg model, we
will first turn to a quantitative discussion using an analytical
method. This method employs the projection operator perturbation theory
(POPT) of Ref.~\onlinecite{PerturbationTheory}, which allows to derive a
rate equation $\partial_t C_{J[S],q}(t) = - \gamma_{J[S],q}(t) \,
C_{J[S],q}(t)$ for the current [density] autocorrelation. This rate
equation gives access to $D_q(t)$ through the right-hand [central]
expression in Eq.~(\ref{D}). The POPT yields a short-time expansion for
the decay rate $\gamma_{J[S],q}(t)$, the terms of which can be evaluated
from a decomposition $H=H_0+H_1$, if the observable of the autocorrelation
$C_{J[S],q}(t)$ is a conserved quantity for the unperturbed Hamiltonian
$H_0$. For the current, we choose the XY model for $H_0$, in which $J_q$
is conserved {\it only} at $q=0$. For the density, we choose the Ising
model for $H_0$, in which $S_q$ is conserved for {\it all} $q$. Then for
short times we obtain approximately:
\begin{eqnarray}
\frac{\gamma_{J,0}(t)}{\Delta J} &\approx& \frac{\Delta J t}{2} +
\frac{(\Delta J t)^3}{24} + {\cal O}[(\Delta J t)^5], \,\, t J
\lesssim 1.5 \quad \label{gJ} \\
\frac{\gamma_{S,q}(t)}{\tilde{q} J} &\approx& \frac{\tilde{q} J
t}{2} + \frac{(\tilde{q} J t)^3}{16} + {\cal O}[(\tilde{q} J t)^5],
\,\, t J\lesssim \frac{2.1}{\Delta} \quad \label{gS}
\end{eqnarray}
For the full quantitative evaluation of $D_q(t)$ we determine the
leading-order term in Eqs.~(\ref{gJ}) and (\ref{gS}) numerically exact,
following the scheme in Ref.~\onlinecite{PerturbationTheory}, which leads
to small changes only. We note that for a complete integration of
Eq.~(\ref{D}) the high-temperature limits of the static correlation
functions are needed, i.e.~$C_{J[S],q}(0) = 1/8$ $[1/4]$.

For $q=0$, we obtain from the POPT and the right-hand expression
in Eq.~(\ref{D}) a leading order prediction as follows: the current
autocorrelation $C_{J,0}(t)$ decays even {\it stronger} than in a
simple relaxation-time approximation, i.e.~according to a Gaussian,
and the diffusion coefficient $D_0(t)$ is an error function. This
prediction is consistent with using Eq.~(\ref{gJ}) for {\it all} times,
which is justified because the current relaxation time from Eq.~(\ref{gJ})
is $t_0 J \approx 1.9/\Delta$, $C_{J,0}(t_0)/C_{J,0}(0)= 1/e$. Therefore
for $\Delta=1.5$ or $2.0$, as in this Letter, $D_0(t)$ has saturated for
times within $t J \lesssim 1.5$. The resulting quantitative $D_0(t)$ is
shown in Fig.~\ref{Fig1} (red/dark-colored, solid curves): Here, $D_0(t)$
first increases linearly but then saturates at a constant value $D_0 J
\approx 0.88/\Delta$, which is reached at $t J \gtrsim \tau_0 J \approx
3.0/\Delta$. For the remainder of this Letter we refer to the saturation
time $\tau_q$ as the `current relaxation time', rather than $t_q$, since
it can be extracted more precisely from later numerical data. We emphasize
that our value of $D_0$ agrees remarkably well with other approaches in
Refs.~\onlinecite{prelovsek2004, DiffusionConstantLindblad}. It is worth
to mention that, for $\Delta \rightarrow \infty$, the $1/\Delta$-scaling
of $D_0$ may break down due to possible recurrences of $C_{J,0}(t)$, see
also Ref.~\onlinecite{znidaric2011} for an alternative point of view.

For $q \neq 0$, we obtain from the POPT and the central expression
in Eq.~(\ref{D}) a prediction for the full momentum-dependence of
$D_q(t)$. This prediction is only valid at short times, set by
Eq.~(\ref{gS}). The resulting quantitative $D_q(t)$ is depicted in
Fig.~\ref{Fig1} (green/light-colored, solid curves). Clearly, $D_q(t)$
is {\it not} constant in the short-time domain as a function of $q$.
The $q$-dependence arises from the next-to-leading order term of the
POPT and becomes significant for momenta above $q \sim 0.2 \pi \,
\Delta$, at $\Delta = 1.5$ and $2.0$, and is particularly evident
for $q=\pi$.

\begin{figure}[t]
\includegraphics[width=0.85\columnwidth]{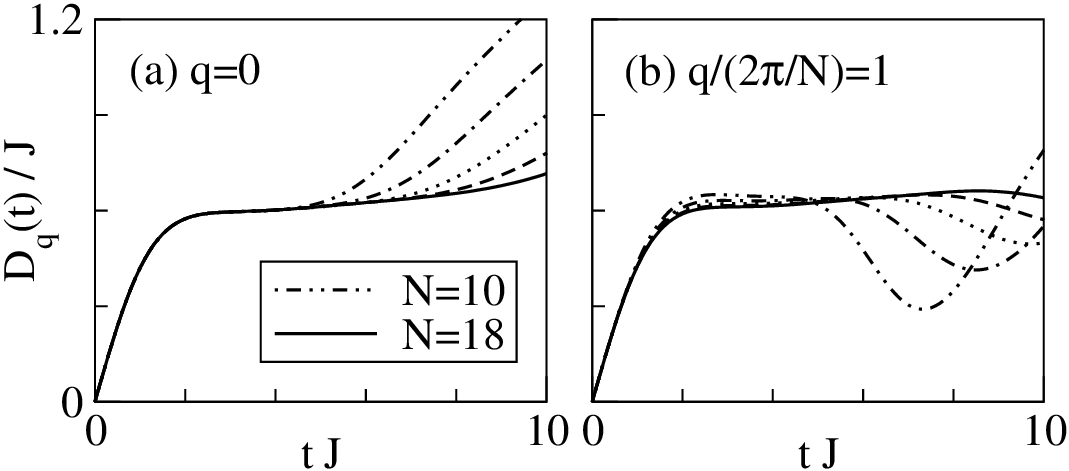}
\caption{(color online) Finite-size scaling results for the diffusion
coefficient $D_q(t)$ at (a) $q=0$, (b) $q/(2\pi/N)=1$ for different $N=10$,
$12$, $\ldots$, $18$ at $\beta = 0$ and $\Delta =1.5$. In (a) finite-size
variations can be neglected for $t J \lesssim 10$ at $N=18$. While
$q = \mathrm{const.}$ cannot be maintained in (b), the tendency is similar
to (a).} \label{Fig2}
\end{figure}

\begin{figure}[b]
\includegraphics[width=0.85\columnwidth]{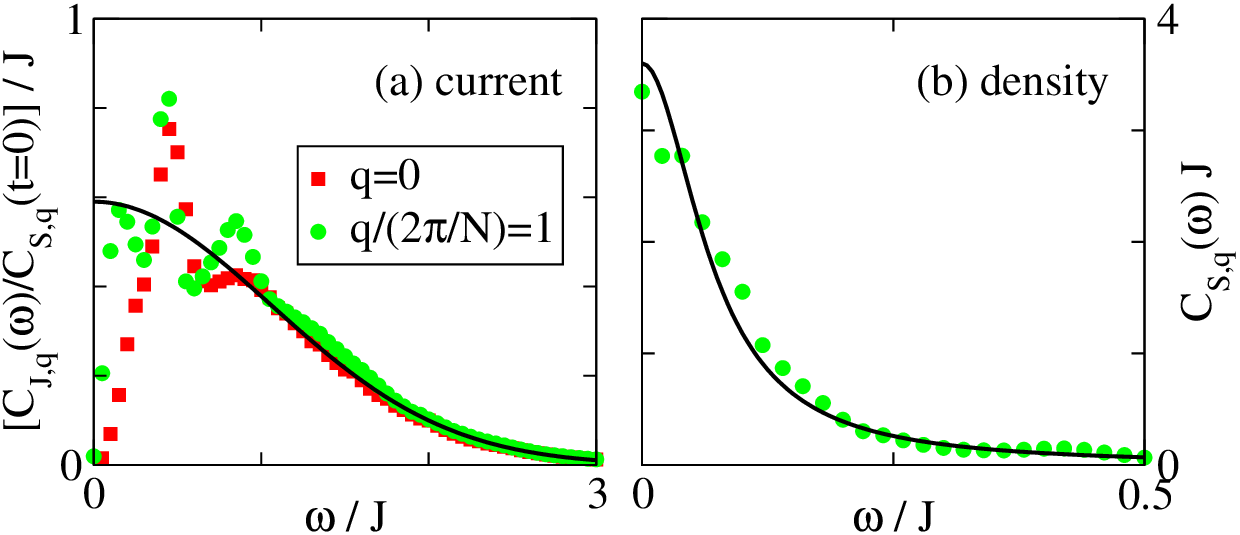}
\caption{(color online) Spectrum of the (a) current and (b) density
autocorrelation at $\beta = 0$ and $\Delta = 1.5$. ED results are
shown for $N= 18$ (symbols). In (a) a Gaussian with height $D_0$ and
in (b) a Lorentzian with width $\tilde{q}^2 \, D_0$ are indicated
for comparison (curves), using $D_0 J \approx 0.59$ from the
POPT.} \label{Fig3}
\end{figure}

In order to complete the picture at $\beta = 0$ for arbitrary momenta and
times we apply exact diagonalization (ED) to chains of length $N = 18$,
allowing for a $q$-grid with $\delta q \approx 0.11 \pi$. Figure~\ref{Fig1}
depicts our results for $D_q(t)$. Several comments are in order. First,
Figs.~\ref{Fig1}~(a) and (b) show a convincing agreement between ED and
both POPTs within their respective ranges of validity, which corroborates
our analysis.  Next, a given density mode at wave vector $q$ shows the
signature of a diffusive decay if there is a plateau with $D_q(t) \approx
\mathrm{const.}$ within a `long-time' window $\tau_q \lesssim t \lesssim
t_D$ with $\tau_q \ll t_D$. Clearly, Fig.~\ref{Fig1} shows that the first
three (four) momenta for $\Delta= 1.5$ ($2.0$) feature such
plateaus. Long-time deviations from this behavior can have several origins,
such as finite-size effects, finite Drude weights, or other low-frequency
anomalies. Most remarkable, the plateau values of $D_q (\tau_q\lesssim
t\lesssim t_D)\approx D_0$ and $\tau_q\approx\tau_0$ are {\it independent}
of momentum, to within the typical finite-size variations which occur for
$N=16 \rightarrow 18$. Finite-size effects are further quantified in
Fig.~\ref{Fig2} for the interval $t J \lesssim 10$. While a constant finite
$q$ cannot be maintained as $N$ varies, it is still obvious that system
sizes of $N=18$ are completely sufficient to determine $D_q$ at the
plateau. The weak dependence on $q$ has to be contrasted against the
significant $q$-dependence for larger momenta. In agreement with the POPT
for $q=0$, the values of $D_0$ and $\tau_0$ from ED can be scaled onto a single
expression for the two anisotropies studied, namely, $D_0 \, J \approx
0.88/\Delta$ for $t \, J \gtrsim \tau_0 \, J \approx 3.0/ \Delta$. This is
one of the main results of this Letter, i.e., the existence of an extended
momentum-space region with a $q$-independent diffusion constant $\propto
1/\Delta$.  Clearly, the number of momenta to which the diffusion criterion
applies is smaller in Fig.~\ref{Fig1}~(a) than in (b). For both $\Delta =
1.5$ and $2.0$, we find no indications of diffusion for $q\gtrsim 0.22
\pi\Delta\equiv q_D$. Instead, $D_{q>q_D}(t)$ displays divergent behavior
due to oscillations of $C_{S,q}(t)$ with time, preventing diffusive
behavior to occur.  These oscillations have already been reported in
Ref.~\onlinecite{fabricius1997} for smaller $\Delta$.

We emphasize that ED results for the {\it spectra} $C_{J[S],q}(\omega)$ at
small $q$ versus frequency $\omega$ agree with our interpretation from the
time domain. E.g., focusing on $\Delta = 1.5$, Fig.~\ref{Fig3}~(a) shows
that the spectrum $C_{J,q}(\omega)/ C_{S,q}(t=0)$ is consistent with a
Gaussian of height $D_0 J \approx 0.59$, as predicted by the POPT at $q=0$.
The low-frequency behavior is still governed by finite-size effects and
deviations from the Gaussian occur at a frequency scale $\omega/J\lesssim
1$, which is independent of $q$. This agrees with the $q$-independent time
scale in Fig.~\ref{Fig2}, where finite-size effects set in. Similar spectra
of $C_{J,q}(\omega)$ have been obtained in Ref.~\onlinecite{mierzejewski2011}
for $q=0$. Note that the (finite size) $q=0$ Drude weight at $\omega = 0$ is
not shown in Fig.~\ref{Fig3}~(a). Figure~\ref{Fig3}~(b) shows that
$C_{S,q}(\omega)$ is consistent with a Lorentzian of width $\tilde{q}^2 D_0$,
again $D_0 J \approx 0.59$, as expected for diffusive density decay.

\begin{figure}[tb]
\includegraphics[width=0.9\columnwidth]{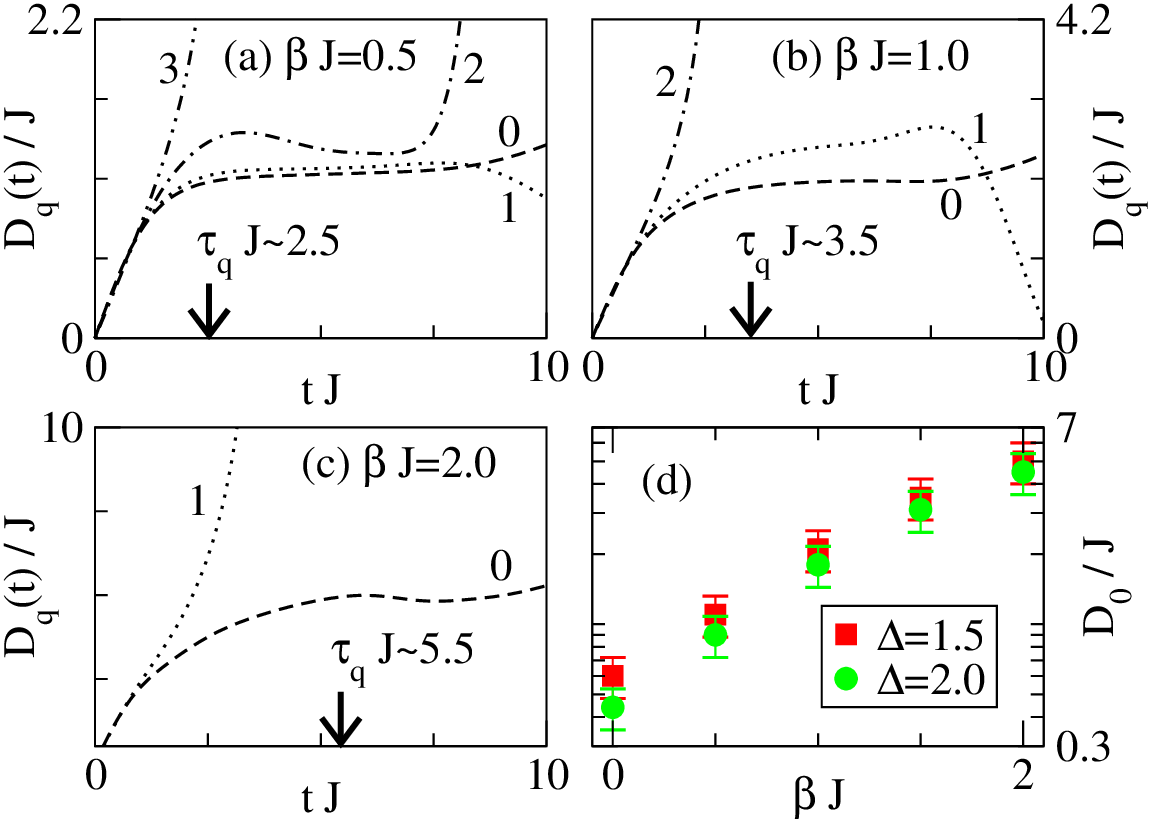}
\caption{ (color online) (a)--(c) ED results for the diffusion
coefficient $D_q(t)$ at different $\beta > 0$ for $\Delta=1.5$ and
$N=18$ (curves). Thick arrows mark the approximate locations of the
current decay time $\tau_q$. (d) The resulting diffusion constant
$D_0$ versus $\beta$ for $\Delta = 1.5$ (squares) and $\Delta = 2.0$
(circles), which can be determined with a precision of $20\%$
(error bars).} \label{Fig4}
\end{figure}

Now we turn to the effects of temperature by increasing $\beta$ from
$0$ to $\beta J = 2$. Since the POPT is not applicable at $\beta
\neq 0$, we focus on the ED results.  Figure~\ref{Fig4}~(a)--(c)
summarizes our findings for $\Delta = 1.5$ and $\beta J = 0.5$, $1$,
$2$. We observe two effects.  First, as the temperature is lowered,
the number of momenta with diffusive density dynamics decreases. At
$\beta J= 0.5$ and $1$ the mode with $q = 0.11 \pi$ still decays
diffusively but for $\beta J = 2$ only the $q = 0$ mode displays
diffusion. Second, as the temperature is lowered, $D_q$ and $\tau_q$
increase significantly. For $q=0$ this increase can be followed up
to $\beta J = 2$. Fig.~\ref{Fig4}~(d) displays $D_0$ versus $\beta$
in a semi-logarithmic plot for $\Delta = 1.5$ and $2.0$. From this
plot, one might be tempted to speculate on an exponential increase
of $D_0$ with $\beta$ beyond the temperature window depicted, see a
related claim in Ref.~\onlinecite{znidaric2011}. However, in view of
the hydrodynamic relation $D_0 = t_0 \, C_{J,0}(0)/C_{S,0}(0)$ this
is a subtle issue. From our numerical analysis, we find $C_{S,0}(0)$
to be the dominant source of $D_0$'s $T$-dependence for $0<\beta
J<2$. But $C_{S,0}(0)$ is not $\propto\exp(c\beta)$ for all $\beta$
\cite{XXZSusceptibility}. An exponential increase of $D_0$ must
further break down as $\beta\rightarrow\infty$ due to the finite
spin gap for $\Delta>1$. We also mention that, for $\Delta=1$ and
$\beta J\gg 1$, the dominant $T$-dependence of $D_0$ stems from
$t_0\propto1/(T \ln T)$ \cite{sirker2009,grossjohann2010}, which is
not exponential.

\begin{figure}[tb]
\includegraphics[width=0.85\columnwidth]{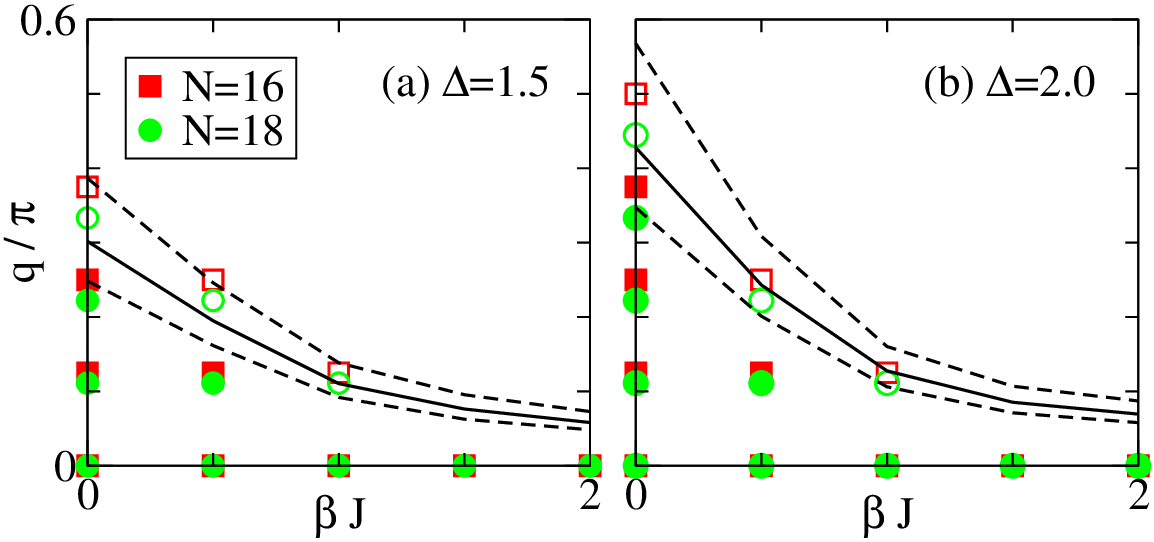}
\caption{(color online) The diffusive range of momenta $q$ versus
$\beta$ for (a) $\Delta = 1.5$ and (b) $\Delta = 2.0$. ED results
are shown for $N=16$ (squares) and $N=18$ (circles). [Open symbols
are borderline values, see Fig.~\ref{Fig1} and
Fig.~\ref{Fig4}~(a)--(c).] The curve $\tilde{q}^2 \, D_0 \, \tau_0 =
1$ is indicated for comparison (solid curves). (Dashed curves
estimate errors for $D_0$ and $\tau_0$.)} \label{Fig5}
\end{figure}

Finally, we turn to a more detailed discussion of the temperature
dependence of the critical momentum $q_D$. To this end, we first
collect all momenta $q\lesssim q_D$ in Fig.~\ref{Fig5}. Then, to
rationalize this, we invoke the standard hydrodynamics criterion
that the relaxation time $1/(\tilde{q}^2 D_q)$ of a diffusive
density mode should be larger than the decay time $\tau_q$ of the
current, or equivalently, that a diffusive density spectrum should
be narrower than the current spectrum, see Fig.~\ref{Fig3}.
Therefore, breakdown of diffusion occurs at $\tilde{q}^2 D_q \,
\tau_q \sim 1$, where we may set $D_q = D_0$ and $\tau_q = \tau_0$,
due to the weak $q$-dependence of these quantities in our case.
Based on our ED results for $D_0$ and $\tau_0$, Fig.~\ref{Fig5}
displays the lines $\tilde{q}^2 D_0 \, \tau_0 = 1$ versus $\beta$
for both $\Delta = 1.5$ and $2.0$ (solid curves). The obvious
agreement between these lines and the boundaries for the collected
values of $q\lesssim q_D$ is a convincing consistency check of our
approach. Apparently, as $\beta$ increases, $q_D$ decreases. In view
of the temperature dependence of $D_0$ and $\tau_0$, this decrease
is also approximately exponential for $0 \leq \beta J \lesssim 2$.
To asses the relevance of finite size effects, Fig.~\ref{Fig5}
contains a comparison between the lines $\tilde{q}^2 D_0 \tau_0 = 1$
and the observed diffusive modes for $N=16$ and $18$ (symbols).
Given the limited resolution of the $q$-grid, the agreement with
these two system sizes is remarkably good.

In summary we have investigated magnetization transport in the
spin-$1/2$ XXZ chain above the isotropic point at finite temperature
and momentum. We found an extended momentum-space region of
spin-diffusion with an approximately time and momentum independent
diffusion constant. The diffusion cut-off wave vector (diffusion
constant) was found to scale approximately linear with the (inverse)
anisotropy and to decrease (increase) strongly with the inverse
temperature.

This work was supported by the Deutsche Forschungsgemeinschaft
through FOR912, Grants No.~BR 1084/6-1 and 1084/6-2, and the
European Commission through MC-ITN LOTHERM, Grant
No.~PITN-GA-2009-238475. WB thanks the `Platform for
Superconductivity and Magnetism', Dresden, for kind hospitality.

\end{document}